\documentclass[%
 reprint,
superscriptaddress,
showpacs,preprintnumbers,showkeys,
 amsmath,amssymb,
 aps,
pra,
]{revtex4-1}
\usepackage{float}
\usepackage{graphicx}
\usepackage{dcolumn}
\usepackage{bm}
\usepackage{epstopdf}

\usepackage{hyperref}

\begin{document}


\title{Transverse spin in scattering of focused radially and azimuthally polarized  vector beams}
%
\author{Ankit Kumar Singh}
\affiliation{Indian Institute of Science Education and Research Kolkata, Mohanpur 741246}
\author{Sudipta Saha}
\email{sudipta.20a@gmail.com}
\affiliation{Indian Institute of Science Education and Research Kolkata, Mohanpur 741246}
\author{Subhasish Dutta Gupta}
\affiliation{School of Physics, University of Hyderabad, Hyderabad 500046, India}
\author{Nirmalya Ghosh}
\affiliation{Indian Institute of Science Education and Research Kolkata, Mohanpur 741246}


\begin{abstract}

We study the effect of focusing of the radially and azimuthally polarized vector beams on the spin angular momentum (SAM) density and Poynting vector of scattered waves from a Mie particle. Remarkably, the study reveals that the SAM density of the scattered field is solely transverse in nature for radially and azimuthally polarized incident vector beams; however the Poynting vector shows usual longitudinal character. We also demonstrate that the transverse SAM density can further be tuned with wavelength and focusing of the incident beam by exploiting the interference of different scattering modes. These results may stimulate further experimental technique of detecting the transverse spin and Belinfante\textquoteright s spin-momentum densities.


\end{abstract}

\maketitle


The linear and the angular momenta are the main dynamical properties of electromagnetic waves, which play important roles in various light-matter interactions\cite{allenbook,ngbook,cohen1992atom}.  The electromagnetic momentum density can also be represented by Poynting linear momemtum density ($\mathbf{P}$). In case of ideal plane waves, the Poynting vector has only longitudinal component along the direction of wave propagation specified by the wave-vector \textbf{k}. However, for inhomogeneous optical fields, the direction of the Poynting vector does not necessarily coincide with the direction of \textbf{k}\cite{allenbook,ngbook}. In fact, it has been shown that the Poynting vector \textbf{P} contains contributions of the orbital ($\mathbf{P_o}$) and the spin momenta ($\mathbf{P_S}$)\cite{Bliokh2014,bliokh_phyrep,ngbook,allenbook}. The former (orbital $\mathbf{P_o}$) is the canonical momentum associated with the wave vector \textbf{k} and is responsible for exerting radiation pressure force. The solenoidal spin momentum ($\mathbf{P_S}$), on the other hand, has long been known as a virtual entity in a sense that it does not exert any optical  pressure in the dipole approximation and is only responsible for generating the spin angular momentum (SAM, $\mathbf{S}$) ($\mathbf{P_S}=\dfrac{1}{2}\bigtriangledown\times \mathbf{S}$)\cite{Belinfante_1940,natphys_2016}. Usually, SAM (\textbf{S}) has only longitudinal component (along \textbf{k}) and is associated with the circular/elliptical polarization of light waves with helicity $\sigma$ in the range $-1\leq\sigma\leq+1$. Thus, the spin momentum ($\mathbf{P_S}$) is directed orthogonal to \textbf{k} and depends upon the helicity (degree of circular polarization) of the wave (opposite for $\sigma=\pm1$)\cite{ngbook}. However recent studies have shown that apart from the longitudinal SAM  ($\mathbf{S}$) and momentum  ($\mathbf{P}$) densities, structured optical fields (e.g. evanescent field, surface plasmon polaritons) having phase shifted longitudinal field components may lead to helicity-independent transverse SAM and helicity-dependent transverse momentum densities  \cite{Aiello_2015,measuring_2015,Bliokh2014,bliokh2015quantum,spinsurface_2012,Bauer_2016}. These quantities are proved to be useful in the context of studying spin-optical effects in nano-optical systems e.g. for realizing the spin-momentum locking (the resulting optical analog of quantum Hall effect), waveguided plasmonic antennas etc\cite{bliokh2015quantum,application1,application2,application3}.  \\

Indeed the transverse momentum and SAM densities have been successfully observed and measured, using Mie resonant probe particles \cite{measuring_2015,natphys_2016,Bliokh2014}. Recently it has been shown that the scattering from a Mie particle can itself show transverse SAM and transverse momentum components for plane wave excitation \cite{Saha_2016,mdm}. However the strength as well as the spatial extent of these are limited to very near field of the particle and mainly dominated by the longitudinal component of them, making it a bit challenging to detect experimentally. Therefore it is desirable to find out systems and novel means where SAM density is only transverse in nature with considerably larger spatial extent. \\

Here we report that indeed radially and azimuthally polarized vector beams provide an interesting avenue towards it. Our theoretical calculation has shown that the complex source focused vector beams \cite{orlov2010,orlov2012} possess only transverse component of SAM density.
Also, the structure of field scattered from a spherical particle for incident radially and azimuthally polarized light enables one to study the transverse spin of transverse electric (TE) and transverse magnetic (TM) modes of the sphere separately. For radially polarized incident beams, the electric field of the scattered waves has only contribution from TM ($a_n$) scattering modes; while for the azimuthally polarized light, the contribution comes from TE ($b_n$) scattering modes only. Therefore we study the transverse SAM in the scattering of radially and azimuthally polarized light from a (a) plasmonic sphere, (b) magnetic mode resonant Silicon sphere  and (c) dielectric micro-sphere. A brief study about effect of focusing on the transverse SAM and momentum densities of the field scattered from the sphere for incident focused beam is also discussed. \\

\begin{figure}[ht]
	\centering
	\includegraphics[width=0.8\linewidth,keepaspectratio]{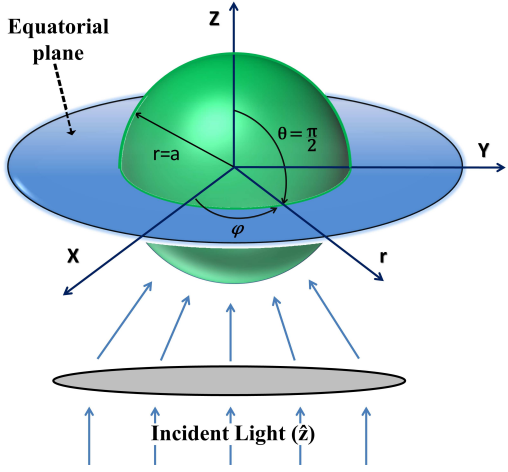}
	\caption{\label{fig1}Schematic view of the scattering geometry: the scatterer is placed at the center of the Cartesian coordinate (XYZ) system and the particle is illuminated by a focused beam propagating along Z-direction. The SAM density and the Poynting vector are studied at the equatorial plane (shown by the blue color).}
\end{figure}

Fig. \ref{fig1} shows the schematic of scattering geometry. Here, the focused field is propagating along Z direction and the polarization axes in laboratory frame is given by the two orthogonal axes X and Y. The scattering angles $\theta$ and $\phi$ are assigned with respect to Z and X axes respectively, as shown in the figure.\\

The field scattered by a homogeneous spherical particle of radius \textit{a} kept in a non-absorbing medium of permittivity $\epsilon$  and permeability $\mu$ for a highly focused beam of frequency $\omega$ can be obtained using Mie-like theory for highly focused complex source beam. The expressions for scattered electric ($\mathbf{E_s}$) and magnetic ($\mathbf{H_s}$) fields for focused light of arbitrary incident polarization are given by \cite{orlov2012}:
\begin{eqnarray}
\mathbf{E_{s}}= \sum\limits_{n=1}^\infty\sum\limits_{m=-n}^n(a_nA_{mn} {\mathbf{\tilde N}}^{(3)}_{mn}+b_n B_{mn}{\mathbf{\tilde M}}^{(3)}_{mn} ) \\
\mathbf{H_{s}}=\dfrac{-ik}{\omega \mu} \sum\limits_{n=1}^\infty\sum\limits_{m=-n}^n(a_nA_{mn} {\mathbf{\tilde M}}^{(3)}_{mn}+b_n B_{mn}{\mathbf{\tilde N}}^{(3)}_{mn})
\end{eqnarray}
Here, $a_n$ and $b_n$ are the classical Mie coefficients for TM and TE scattering modes; $\mathbf{\tilde M}^{(3)}$ and $\mathbf{\tilde N}^{(3)}$ represent the normalized vector spherical harmonics(VSH) \cite{kongbook}; \textit{A} and \textit{B} decide the amplitude of scattered field depending upon incident polarization and focusing. In case of focused linear polarized light beam parallel to X axis, with beam waist $w_o$ and Rayleigh distance $z_0=kw^2_0/2$, the scattered fields can be simplified in term of VSH ($\mathbf{M}$ and $\mathbf{N}$) as

\begin{eqnarray}
\mathbf{E_{s}}= \sum\limits_{n=1}^\infty (a_n \mu_n {\mathbf{N}}_{e1n}+b_n \beta_n {\mathbf{M}}_{01n} ) \\
\mathbf{H_{s}}=\dfrac{-ik}{\omega \mu} \sum\limits_{n=1}^\infty (a_n\mu_n {\mathbf{M}}_{e1n}+b_n \beta_n {\mathbf{N}}_{o1n})
\end{eqnarray}
with 
\begin{eqnarray}
\beta_n=iU_0\Big[\dfrac{(2n+1) j^*_n(ikz_0)}{n(n+1)}-\Big( \dfrac{j^*_{n-1}(ikz_0)}{n}- \dfrac{j^*_{n+1}(ikz_0)}{n+1} \Big)\Big] \nonumber \\
\mu_n=iU_0\Big[\dfrac{(2n+1) j^*_n(ikz_0)}{n(n+1)}+\Big( \dfrac{j^*_{n-1}(ikz_0)}{n}- \dfrac{j^*_{n+1}(ikz_0)}{n+1} \Big)\Big] \nonumber
\end{eqnarray}
where $j_n^*$ is the spherical Bessel function and the normalization constant $U_o=kz_o/\sinh(kz_0)$.\\

The expression of the fields can then be used to calculate the normalized SAM density (\textbf{S}) and the Poynting vector (\textbf{P}) of scattered wave as\cite{bliokh2017optical,bliokh2017opticalprl,mdm,Saha_2016}:
\begin{eqnarray}
\mathbf{S}&=&\dfrac{Im(\tilde{\epsilon} \mathbf{E^*} \times \mathbf{E}+\tilde{\mu} \mathbf{H^*} \times \mathbf{H})}{\omega (\tilde{\epsilon} |\mathbf{E}|^2+\tilde{\mu} |\mathbf{H}|^2)} \label{eq1} \\
\mathbf{P}&=&\dfrac{1}{2} Re(\mathbf{E}\times\mathbf{H}^*) \label{eq2}
\end{eqnarray}
with $\tilde{\epsilon}=\epsilon+\omega \ d\epsilon/d\omega$ and $\tilde{\mu}=\mu+\omega \ d\mu/d\omega$ for dispersive medium. However for non-dispersive medium, these are given as $\tilde{\epsilon}=\epsilon$ and $\tilde{\mu}=\mu$ with $\epsilon$ and $\mu$ being the permittivity and permeability respectively.
\begin{figure}[ht]
	\centering
	\includegraphics[width=\linewidth]{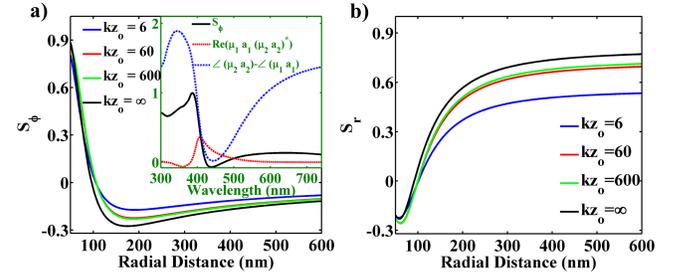}
	\caption{\label{fig2} Computed radial dependence of the (a) transverse ($S_\phi$) and (b) longitudinal ($S_r$) components of SAM ($\mathbf{S}$) density of scattered wave at equatorial plane for a silver (Ag) sphere ($a=50nm$) kept in water for an incident focused right circular polarized light of wavelength $\lambda\approx380nm$ with different beam waists . The inset of (a) shows the wavelength dependence of strength of interference (red dotted) of effective dipolar ($\mu_1 a_1$) and quadrupolar ($\mu_2 a_2$) modes, their phase difference (blue dotted, in radians) and $S_\phi$  at a radial distance $r=50 nm$. }
\end{figure}

The above expression along with the symmetry of the scattering geometry can be used to obtain the normalized SAM density $\mathbf{S}$ and Poynting vector $\mathbf{P}$ of the scattered field for incident focused right/left circular polarized (RCP/LCP) light. The components of $\mathbf{S}$ and $\mathbf{P}$ for focused RCP/LCP beams follow 
the same symmetry relations as is observed in case of scattering of plane waves from spherical scatters \cite{Saha_2016,mdm}. Fig. \ref{fig2} shows computed normalized transverse and longitudinal SAM densities for different beam spot sizes (focusing) for incident right circular polarization (RCP) for a silver (Ag) nanosphere of radius $a=50$ nm kept in water (refractive index $1.33$) at wavelength $\lambda \approx 380$ nm \cite{Soni_2014}. It can be seen that with decrease of beam waist $w_o$ (increasing focusing) a considerable decrease in the longitudinal component of SAM density (Fig. \ref{fig2}(b)) is found. However, the magnitude of transverse component ($S_\phi$) of SAM density for incident circular polarization is not affected much (Fig. \ref{fig2}(a)). Thus the effective transverse SAM density $S_\phi$ at the equatorial plane is observed to be enhanced in the far field of the scatterer. The inset of Fig. \ref{fig2}(a) shows wavelength variation of the transverse SAM density $S_\phi$ at the surface of the scatterer for an incident focused RCP beam characterized by $kz_o=600$. From the inset of Fig.\ref{fig2}(a), it is observed that at the wavelength $\lambda \approx 380$nm, the transverse SAM density  $S_\phi$ gets enhanced because of the constructive interference of the effective electric dipolar ($\mu_1 a_1$) and electric quadrapolar ($\mu_2 a_2$) modes of the sphere. Although focusing does increase the effective contribution of $S_\phi$ however the longitudinal component still dominates over it, except for the extreme near field. Therefore, a system is desirable with no or very small contribution of longitudinal component of SAM density. Indeed, the vector (radially and azimuthally polarized) beams serve the purpose as will be shown in following discussion. \\

\begin{figure}[ht]
	\centering
	\includegraphics[width=\linewidth]{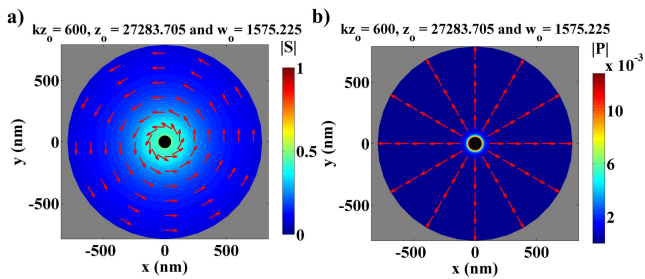}
	\caption{\label{fig3} The computed (a) SAM ($\mathbf{S}$) density and (b) Poynting ($\mathbf{P}$) vector of scattered wave at the equatorial plane from a plasmonic silver (Ag) sphere (a=50) kept in water, for incident focused radially polarized beam with beam spot $w_o\approx1575$. }
\end{figure}
The scattered fields for incident radial ($E_s^{rad}/H_s^{rad}$) and azimuthal ($E_s^{azi}/H_s^{azi}$) polarizations are given as \cite{orlov2010}:
\begin{eqnarray}
\mathbf{E}_s^{rad}=- \sum\limits_{n=1}^\infty a_n \alpha_n {\mathbf{N}}_{e0n}; \mathbf{H}_s^{rad}=\dfrac{ik}{\omega \mu} \sum\limits_{n=1}^\infty a_n \alpha_n {\mathbf{M}}_{e0n} \\ \mathbf{E}_s^{azi}=- \sum\limits_{n=1}^\infty b_n \alpha_n {\mathbf{M}}_{e0n}; \mathbf{H}_s^{azi}=\dfrac{ik}{\omega \mu} \sum\limits_{n=1}^\infty b_n \alpha_n {\mathbf{N}}_{e0n}
\end{eqnarray}
where
\begin{equation}
\alpha_n=\dfrac{kz_0(2n+1)}{sinh(kz_0)}j_n(ikz_0) \nonumber
\end{equation}
A few things must be noted from the above expressions: (a) depending on the input polarization either TM or TE ($a_n$ or $b_n$) will contribute in scattering from the particle, (b) SAM density will only have contributions from the TM ($a_n$) modes for incident radial polarization and TE ($b_n$) modes for azimuthal polarization and, (c) the scattered field will always have SAM density and Poynting vector which are solely  transverse and longitudinal to the direction of propagation respectively.\\

Fig.\ref{fig3} shows the calculated SAM density and Poynting vector for light scattered from the same Ag nanosphere for incident radially polarized light of wavelength $\lambda\approx380$ nm. This wavelength is chosen as the interference of effective electric dipolar ($\alpha_1 a_1$) and electric quadrapolar ($\alpha_2 a_2$) modes gives a wavelength variation for input radially polarized light similar to the one shown in Fig. \ref{fig2}(a) (data not shown here). The perfect transverse and longitudinal nature of the SAM density and Poynting vector are evident respectively from Fig.\ref{fig3} . There is not much effect of the beam waist on the transverse SAM unless the limit of $kz_o\sim n$ is probed. However, a considerable decrease in Poynting vector strength at equatorial plane was observed (not shown) due to decrease in expansion coefficients ($\alpha_n$) with decreasing beam waist of the focused beam.

\begin{figure}[ht]
\centering
\includegraphics[width=\linewidth]{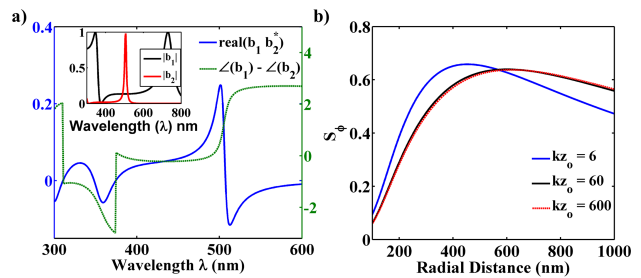}
\caption{\label{fig4}(a) The strength of interference (blue, left axis) of magnetic dipolar ($b_1$) and magnetic quadrapolar ($b_2$) modes and the phase difference (green, right axis) between the modes observed at a radial distance r=100 nm. The inset shows contribution of various modes in scattering cross-section of the sphere (b) The radial dependence of the transverse component of spin ($S_\phi$) is shown with different beam waist parameters.}
\end{figure}

\begin{figure*}[ht]
	\includegraphics[width=\linewidth,keepaspectratio]{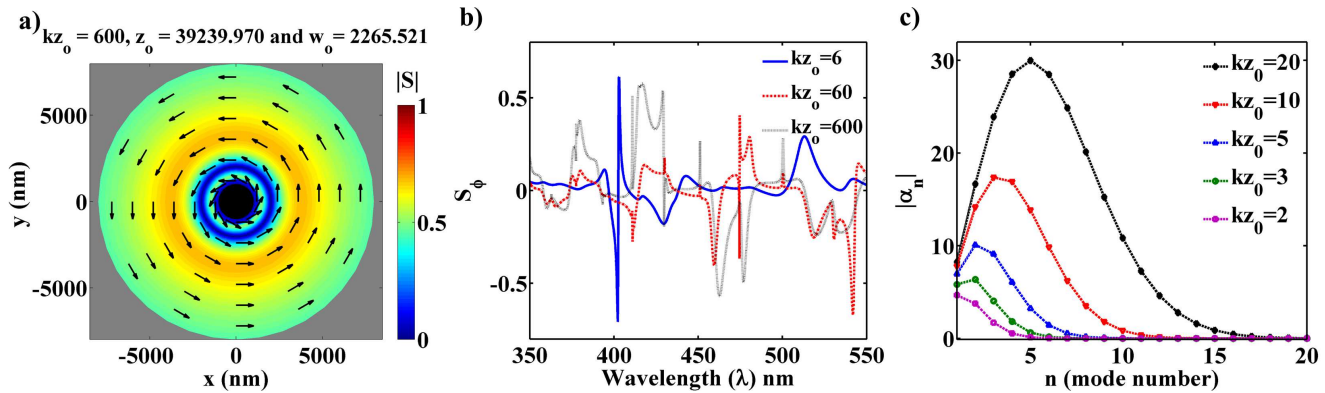}
	\caption{ \label{fig5} The calculated SAM density distribution of the field scattered by a spherical dielectric scatterer ($a=1\mu m$ and refractive index$=1.6$) for azimuthal polarized incident field, in case (a)$ \lambda=410$ nm  (b) The plot showing spectral variation of transverse spin ($S_\phi$) for the dielectric scatterer with azimuthal incidence and it’s tuning with the collimation distance or focusing parameter; (c) Dependence of the expansion coefficients ($\alpha_n$) on the order of multipole for different collimation distance (k$z_0$).}
\end{figure*}

It is to be noted that the magnetic modes are not excited in the plasmonic nanosphere due to small size parameter of the scatterer, we thus choose a Silcon (Si) sphere of radius $100$ nm having a refractive index 3.5 kept in air to study the scattering of incident focused azimuthal polarized light from the contribution of lower order modes only \cite{sisphere,magneticl8}. Here, transverse SAM density ($S_\phi$) gets enhanced due to constructive interference of lower order TE modes ($b_1$ and $b_2$ modes) for incident azimuthal polarization as shown in Fig. \ref{fig4} and the interference maxima occurs at $\lambda \approx 500$nm corresponding to the enhanced transverse spin at the wavelength. The scattering of azimuthal polarized light from the Si sphere shows similar trends as the scattered field from plasmonic nanosphere incident by radially polarized focused beam.\\

Now, we study the effect of interference of multiple TM ($a_n$) or multiple TE ($b_n$) modes on transverse SAM density when excited simultaneously. A dielectric sphere of radius $1\mu$m and refractive index 1.6 kept in air was chosen to observe the multiple modes interference effect. Fig. \ref{fig5}(a) shows the SAM density observed in the scattered field for incident azimuthal polarization on the dielectric scatterer at $\lambda\approx410 nm$. Fig. \ref{fig5}(a) shows that the interference effect could lead to huge enhancement in the amplitude as well as the spatial extent ($\sim$ a \textit{few micron}) of the transverse SAM in the scattered field. It can be seen from Fig. \ref{fig5}(b) that transverse SAM density ($S_\phi$) of the scattered field can be tuned with the wavelength of incident light for selective mode excitation, further such spectral variation of transverse spin can be tuned with the collimation distance or focusing parameter. The dependence of expansion coefficient ($\alpha_n$) on collimation distance ($kz_o$) is shown in Fig. \ref{fig5}(c). It can also be seen from the figure that collimation distance tunes the expansion coefficient ($\alpha_n$) of various modes (n) leading to spectral tunability of $S_\phi$ with collimation distance as seen in Fig. \ref{fig5}(b). Fig. \ref{fig5}(c) also demonstrates that for small collimation distance (i.e. tight focusing) higher order modes cannot be excited and  only lower order modes participate in the scattering from the particle.\\

In conclusion we have shown the effect of focusing on transverse SAM of the scattered field from a Mie particle. It is shown that for the incident complex source radial or azimuthal beams, the field scattered from Mie scatterer shows completely transverse SAM density with a Poynting vector having the usual longitudinal character. The results show that the spatial extent of the transverse SAM can be tuned with wavelengths and focusing of the incident beam because of the modulated interference of various scattering modes contributing in the scattered field. These findings may lead to further experimental studies for detecting the the transverse spin and Belinfante's spin momentum densities.\\

Authors acknowledge the Indian Institute of Science Education and Research, Kolkata for the funding and facilities. AKS acknowledges the Council of Scientific and Industrial Research (CSIR), Govt. of India for research fellowship.

\bibliography{radialref.bib}

\end{document}